\documentclass[11pt]{article}
\usepackage{hyperref}
\hypersetup{
pdftitle={Blurring Out Cosmic Puzzles},
pdfauthor={Yann Ben\'etreau-Dupin},
colorlinks=true,
linkcolor={black},
citecolor={blue},
urlcolor={blue},
}

\usepackage{lastpage,sectsty,setspace,commath,amsfonts,natbib,fancyhdr,wrapfig,graphicx,array,chngpage}

\newcommand{\ud}{\,\mathrm{d}} 

\usepackage{footmisc}

\usepackage{relsize}

\author{Yann Ben\'etreau-Dupin\footnote{Department of Philosophy \& Rotman Institute of Philosophy, Western University, Canada, \href{mailto:ybenetre@uwo.ca}{ybenetre@uwo.ca}} \thanks{Forthcoming in \emph{Philosophy of Science} (PSA 2014). This paper stems from discussions at the UCSC-Rutgers Institute for the Philosophy of Cosmology in the summer of 2013. I am grateful to Chris Smeenk and Wayne Myrvold for discussions and comments on earlier drafts. For helpful discussions, I thank participants at the Imprecise Probabilities in Statistics and Philosophy workshop at the Munich Center for Mathematical Philosophy in June of 2014 and at the Graduate Colloquium series at Western University.
}}

\title{Blurring Out Cosmic Puzzles}

\begin{document} 
 
\maketitle
\thispagestyle{empty}

\begin{abstract}
The Doomsday argument and anthropic reasoning are two puzzling examples of probabilistic confirmation. In both cases, a lack of knowledge apparently yields surprising conclusions.  Since they are formulated within a Bayesian framework, they constitute a challenge to Bayesianism. Several attempts, some successful, have been made to avoid these conclusions, but some versions of these arguments cannot be dissolved within the framework of orthodox Bayesianism. I show that adopting an imprecise framework of probabilistic reasoning allows for a more adequate representation of ignorance in Bayesian reasoning and explains away these puzzles.
\end{abstract}


\section{Introduction}

The Doomsday paradox and the appeal to anthropic bounds to solve the cosmological constant problem are two examples of puzzles of probabilistic confirmation. These arguments both make `cosmic' predictions: the former gives us a probable end date for humanity, and the second a probable value of the vacuum energy density of the universe. They both seem to allow one to draw unwarranted conclusions from a \emph{lack of knowledge}, and yet one way of formulating them makes them a straightforward application of Bayesianism. They call for a framework of inductive logic that allows one to represent ignorance better than what can be achieved by orthodox Bayesianism so as to block these conclusions.

\subsection{The Doomsday Argument}
\label{Doomsday}

The Doomsday argument is a family of arguments about humanity's likely survival.\footnote{See, e.g., \citep[][\S~6--7]{Bostrom2002}, \citep{Richmond2006} for reviews.} There are mainly two versions of the argument discussed in the literature, both of which appeal to a form of Copernican principle (or principle of typicality or mediocrity). A first version of the argument endorsed by, e.g., John \citet{Leslie1990} dictates a probability shift in favor of theories that predict earlier end dates for our species assuming that we are a typical---rather than atypical---member of that group.

The other main version of the argument, referred to as the `delta-$t$ argument,' was given by Richard \citet{Gott1993} and has provoked both outrage and genuine scientific interest.\footnote{See, e.g., \citep{Mackay1994} for opprobrium and \citep{Wells2009,Griffiths2006} for praise.} It claims to allow one to make a prediction about the total duration of any process of indefinite duration based only on the assumption that the moment of observation is randomly selected. A variant of this argument, which gives equivalent predictions, reasons in terms of random ``sampling'' of one's rank in a sequential process \citep{Gott1994}.\footnote{The latter version doesn't violate the reflection principle---entailed by conditionalization---according to which an agent ought to have now a certain credence in a given proposition if she is certain she will have it at a later time \citep{Monton2001}.} The argument goes as follows:

\par Let $r$ be my birth rank (i.e., I am the $r^{\text{th}}$ human to be born), and $N$ the total number of humans that will ever be born.

\begin{enumerate}
\item Assume that there is nothing special about my rank $r$. Following the principle of indifference, for all $r$, the probability of $r$ conditional on $N$ is $p(r|N)=\dfrac{1}{N}$.
\item Assume the following improper prior probability distribution\footnote{As \citet{Gott1994} recalls, this choice of prior is fairly standard (albeit contentious) in statistical analysis. It's the Jeffreys prior for the unbounded parameter $N$, such that $p(N)\ud N\propto\ud \ln N\propto\dfrac{\ud N}{N}$. This means that the probability for $N$ to be in any logarithmic interval is the same. This prior is called improper because it is not normalizable, and it is usually argued that it is justified when it yields a normalizable posterior.}\label{Gott_Jeffreys} for $N$: $p(N)=\dfrac{k}{N}$. $k$ is a normalizing constant, whose value doesn't matter.
\item This choice of distributions $p(r|N)$ and $p(N)$ gives us the prior distribution $p(r)$:
\[
p(r)=\int_{N=r}^{N=\infty}p(r|N)p(N)\ud N=\int_{N=r}^{N=\infty}\dfrac{k}{N^2}\ud N=\dfrac{k}{r}.
\]
\item Then, Bayes's theorem gives us $p(N|r)=\dfrac{p(r|N)\cdot p(N)}{p(r)}=\dfrac{r}{N^2},$ which favors small $N$.
\end{enumerate}

To find an estimate with a confidence $\alpha$, we solve $p(N\leq x|r)=\alpha$ for $x$, with $p(N\leq x|r)=\mathlarger{\int}_r^x p(N|r)\ud N$. Upon learning $r$, we are able to make a prediction about $N$ with a 95\%-level confidence. Here, we have $p(N\leq 20r|r)=0.95$. That is, we have $p(N>20 r|r)<5\%.$

According to that argument, we can make a prediction for $N$ based only on knowing our rank $r$ and on being indifferent about any value $r$ conditional on $N$ may take. This result should strike us as surprising: we shouldn't be able to learn something from nothing! If $N$ is unbounded, an appeal to our typical position shouldn't allow us to make any prediction at all, and yet it does.

\subsection{Anthropic Reasoning in Cosmology}
\label{anthr}

Another probabilistic argument that claims to allow one to make a prediction from a lack of knowledge is commonly used in cosmology, in particular to solve the cosmological constant problem (i.e., explain the value of the vacuum energy density $\rho_V$). This parameter presents physicists with two main problems:\footnote{See \citep{Carroll2000,Sola2013} for an overview of the cosmological constant problem.}

\begin{enumerate}
\item The time coincidence problem: we happen to live at the brief epoch---by cosmological standards---of the universe's history when it is possible to witness the transition from the domination of matter and radiation to vacuum energy ($\rho_M\sim\rho_V$).
\item There is a large discrepancy---of 120 order of magnitudes---between the (very small) observed values of $\rho_V$ and the (very large) values suggested by particle-physics models.
\end{enumerate}

Anthropic selection effects (i.e., our sampling bias as observers existing at a certain time and place and in a universe that must allow the existence of life) have been used to explain both problems. In the absence of satisfying explanations, anthropic selection effects make the coincidence less unexpected, and account for the discrepancy between observations and possible expectations from available theoretical background. But there is no known reason why having $\rho_M\sim\rho_V$ should matter to the advent of life.

Steven Weinberg and his collaborators \citep{Weinberg1987,Weinberg2000,Martel1998a}, among others, proposed anthropic bounds on the possible values of $\rho_V$.
Furthermore, they argued that anthropic considerations may have a stronger, predictive role. The idea is that we should conditionalize the probability of different values of $\rho_V$ on the number of observers they allow: the most likely value of $\rho_V$ is the one that allows for the largest number of galaxies (taken as a proxy for the number of observers).\footnote{This assumption is contentious; see, e.g., \citep{Aguirre2001} for an alternative proposal.} The probability measure for $\rho_V$ is then as follows:
\[
\ud p(\rho_V)=\nu(\rho_V)\cdot p_\star(\rho_V)\ud\rho_V,
\]
where $p_\star(\rho)\ud\rho_V$ is the prior probability distribution, and $\nu(\rho_V)$ the average number of galaxies which form for $\rho_V$.

\par By assuming that there is no known reason why the likelihood of $\rho_V$ should be special at the observed value, and because the allowed range of $\rho_V$ is very far from what we would expect from available theories, Weinberg and his collaborators argued that it is reasonable to assume that the prior probability distribution is constant within the anthropically allowed range, so that $\ud p(\rho_V)$ can be calculated as proportional to $\nu(\rho_V)\ud\rho_V$ \citep[][2]{Weinberg2000}. Weinberg then predicted that the value of $\rho_V$ would be close to the mean value in that range  (assumed to yield the largest number of observers). This ``principle of mediocrity,'' as Alexander \citet{Vilenkin1995} called it, assumes that we are typical observers.

Thus, anthropic considerations not only help establish the prior probability distribution for $\rho_V$ by providing bounds, but they also allow one to make a prediction regarding its observed value. This method has yielded predictions for $\rho_V$ only a few orders of magnitudes apart from the observed value.\footnote{The median value of the distribution obtained by such anthropic prediction is about 20 times the observed value $\rho_V^{\text{obs}}$ \citep{Pogosian2004}.} This improvement---from 120 orders of magnitude to only a few---has been seen by their proponents as vindicating anthropically-based approaches.

\subsection{The Problem: \emph{Ex Nihilo Nihil Fit}}

The Doomsday argument and anthropic reasoning share a similar structure: 1) a uniform prior probability distribution reflects an initial state of ignorance or indifference, and 2) an appeal to typicality or mediocrity is used to make a prediction. This is puzzling: these two assumptions of indifference and typicality are meant to express neutrality, and yet from them alone we seem to be getting a lot of information. But assuming neutrality \emph{alone} should not allow us to learn anything!

If anthropic considerations were only able to provide us with one bound (either lower or upper bound), then the argument used to make a prediction about the vacuum energy density $\rho_V$ would be analogous to Gott's 1993 `delta-$t$ argument': without knowing anything about, say, a parameter's upper bound, a uniform prior probability distribution over all possible ranges and the appeal to typicality of the observed value favors lower values for that parameter.

I will briefly review several approaches taken to dispute the validity of the results obtained from these arguments. We will see that dropping the assumption of typicality isn't enough to avoid these paradoxical conclusions. I will show that, when dealing with events we are completely ignorant or indifferent about, one can use an imprecise, Bayesian-friendly framework that better handles ignorance or indifference.

\section{Typicality, Indifference, Neutrality}

\subsection{How Crucial to Those Arguments Is the Assumption of Typicality?}

The appeal to typicality is central to Gott's `delta-$t$ argument,' Leslie's version of the Doomsday argument, and Weinberg's prediction. This assumption has generated much of the philosophical discussion about the Doomsday argument in particular. Nick \citet{Bostrom2002} offered a challenge to what he calls the Self-Sampling Assumption (SSA), according to which ``one should reason as if one were a random sample from the set of all observers in one's reference class.'' In order to avoid the consequence of the Doomsday argument, Bostrom suggested to adopt what he calls the Self-Indicating Assumption (SIA): ``Given the fact that you exist, you should (other things equal) favor hypotheses according to which many observers exist over hypotheses on which few observers exist.'' \citep{Bostrom2002} But as he noted himself \citep[122-126]{Bostrom2002}, this SIA is not acceptable as a general principle. Indeed, as Dennis \citet{Dieks1992} summarized:

\begin{quote}
Such a principle would entail, e.g., the unpalatable conclusion that armchair philosophizing would suffice for deciding between cosmological models that predict vastly different chances for the development of human civilization.The infinity of the universe would become certain a priori.
\end{quote}

\par The biggest problem with Doomsday-type arguments resting on the SSA is that their conclusion depends on the choice of reference class. What constitutes ``one's reference class'' seems entirely arbitrary or ill-defined: is my reference class that of all humans, mammals, philosophers, etc.? Anthropic predictions can be the object of a similar criticism: the value of the cosmological constant most favorable to the advent of life (as we know it) may not be the same as that most favorable to the existence of intelligent observers, which might be definable indifferent ways.

Relatedly, Radford \citet{Neal2006} argued that conditionalizing on non-indexical information (i.e., all the information at the disposal of the agent formulating the Doomsday argument, including all their memories) reproduces the effects of assuming both SSA and SIA. Conditionalizing on the probability that an observer with all their non-indexical information exists (which is higher for a later Doomsday, and highest if there is no Doomsday at all) blocks the consequence of the Doomsday argument, without invoking such ad hoc principles, and avoids the reference-class problem \citep[see also][]{Dieks1992}.

Although full non-indexical conditioning cancels out the effects of Leslie's Doomsday argument (and, similarly, anthropic predictions), it is not clear that it also allows one to avoid the conclusion of Gott's version of the Doomsday argument. \citet[20]{Neal2006} dismisses Gott's argument because it rests \emph{only} on an ``unsupported'' assumption of typicality. There are indeed no good reasons to endorse typicality a priori \citep[see, e.g., ][]{Hartle2007}. One might then hope that not assuming typicality would suffice to dissolve these cosmic puzzles. Irit \citet{Maor2008} showed for instance that without it, anthropic considerations don't allow one to really make predictions about the cosmological constant, beyond just providing unsurprising boundaries, namely, that the value of the cosmological constant must be such that life is possible.

\par
\begin{figure}[h!]
\begin{center}
\includegraphics[scale=0.4]{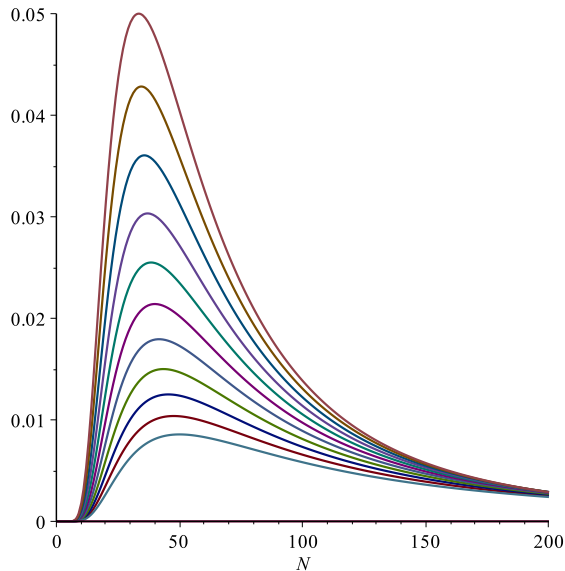}
\caption{Posterior probability distributions for $N$ conditional on $r$, obtained for $r=100$ and assuming different likelihood distributions for $r$ conditional on $N$ (i.e., with different assumptions as to our relative place in humanity's timeline), which each peaks at different values $\tau=\frac{r}{N}$. The lowermost curve corresponds to a likelihood distribution that peaks at $\tau\rightarrow0$, i.e., if we assume $N\rightarrow\infty$.}\label{atypical}
\end{center}
\end{figure}

 My approach in this paper, however, will not be to question the assumption of typicality. Indeed, in Gott's version of the Doomsday argument given in \S~\ref{Doomsday}, we would obtain a prediction \emph{even if we didn't assume typicality}. Instead of assuming a flat probability distribution for our rank $r$ conditional on the total number of humans $N$, $p(r|N)=\dfrac{1}{N}$, let's assume a non-uniform distribution. For instance, let's assume a distribution that favors our being born in humanity's timeline's first decile (i.e., one that peaks around $r=0.1\times N$). We would then obtain a different prediction for $N$ than if we had assumed one that peaks around $r=0.9\times N$. This reasoning, however, yields an unsatisfying result if taken to the limit: if we assume a likelihood probability distribution for $r$ conditional on $N$ sharply peaked at $r=0$, we would \emph{still} obtain a prediction for $N$ upon learning $r$ (see Figure ~\ref{atypical}).\footnote{\citet{Tegmark2005} used a similar reasoning to derive an upper bound on the date of a Doomsday catastrophe.}

Therefore, in Gott's Doomsday argument, we would obtain a prediction at any confidence-level, whatever assumption we make as to our typicality or atypicality, and we would even obtain one if we assume $N\rightarrow\infty$. Consequently, it is toward the question of a probabilistic representation of ignorance or indifference that I will now turn my attention.

\subsection{A Neutral Principle of Indifference?}
\label{neutral}

One could hope that a more adequate prior probability distribution---one that better reflects our ignorance and is normalizable---may prevent the conclusion of these cosmic puzzles (especially Gott's Doomsday argument). The idea that a uniform probability distribution is not a satisfying representation of ignorance is nothing new; this discussion is as old as the principle of indifference itself.\footnote{See, e.g., \citep{Syversveen1998} for a short review on the problem of representing non-informative priors.} As argued by John \citet{Norton2010}, a uniform probability distribution is unable to fulfill invariance requirements that one should expect of a representation of ignorance or indifference:

\begin{list}{-}{}
\item non-additivity,
\item invariance under redescription,
\item invariance under negation: if we are ignorant or indifferent as to whether or not $\alpha$, we must be equally ignorant as to whether or not $\neg\alpha$.\footnote{For an extended discussion about criteria for a representation of ignorance---with imprecise probabilities in particular---see \citep[\S~4--5]{DeCooman2007}.}
\end{list}

For instance, in the case of the cosmological constant problem, if we adopt a uniform probability distribution for the value of the vacuum energy density $\rho_V$ over an anthropically allowed range of length $\mu$, then we are committed to assert, e.g., that $\rho_V$ is 3 times more likely to be found in a any range of length $\dfrac{\mu}{3}$ than in any other range of length $\dfrac{\mu}{9}$. This is very different from indifference or ignorance, hence the requirement of non-additivity for a representation of ignorance.

These criteria for a representation of ignorance or indifference cast doubt on the possibility for a probabilistic logic of induction to overcome these limitations.\footnote{The same goes for improper priors, as was argued, e.g., by \citet{Dawid1973}.} I will argue that an imprecise model of Bayesianism, in which our credences can be fuzzy, will be able to explain away these problems without abandoning Bayesianism altogether.

\section{Dissolving the Puzzles with Imprecise Credence}

\subsection{Imprecise Credence}
\label{IP}

It has been argued \citep[see, e.g.,][]{Levi1974,Walley1991,Joyce2010} that Bayesian credences need not have sharp values, and that there can be imprecise credences (or `imprecise probabilities' by misuse of language). An imprecise credence model recognizes ``that our beliefs should not be any more definitive or unambiguous than the evidence we have for them.'' \citep[320]{Joyce2010}

Joyce defended an imprecise model of Bayesianism in which credences are not represented merely by a range of values, but rather by a \emph{family} of (probabilistic) credence functions. In this imprecise probability model,

\begin{quote}
\begin{enumerate}
\item A believer's overall credal state can be represented by a family $C$ of credence functions [$c_i$] (\ldots). Facts about the person's opinions correspond to properties \emph{common to all the credence functions} in her credal state.
\item If the believer is rational, then every credence function in $C$ is a probability.
\item If a person in credal state $C$ learns that some event $D$ obtains (\ldots), then her post-learning state will be
\[C_D=\left\{c(X|D)=c(X)\dfrac{c(D|X)}{c(D)}: c\in C\right\}.\]

\item A rational decision-maker with credal state $C$ is obliged to prefer one action $\alpha$ to another $\alpha^*$ when $\alpha$'s expected utility exceeds that of $\alpha^*$ relative to \emph{every} credence function in $C$.\label{4} \citep[288, my emphasis]{Joyce2010}
\end{enumerate}
\end{quote}

There are several criteria for decision-making with imprecise probabilities between two propositions. Depending on the criterion chosen, one will prefer an event to another event if
\begin{list}{-}{}
\item it has maximum lower expected value ($\Gamma-$minimax criterion),
\item it has maximum higher expected value ($\Gamma-$maximax),
\item it has maximum expected value for all distributions in the credal set (maximality),
\item it has a higher expected value for at least one distribution in the credal set ($E-$admissibility), or 
\item its lower expected value on all distributions in the credal set is greater than the other event's highest expected value on all distributions (interval dominance).\footnote{This list is not exhaustive, see \citep[\S~8]{Troffaes2007,Augustin2014} for reviews.}
\end{list}

\par This model allows one to simultaneously represent sharp and imprecise credences, but also comparative probabilities. It can accommodate sharp credences and then the usual condition of additivity. But it can also accommodate less sharply defined relationships when credences are fuzzy. It does so by means of a family of credence functions, each of which is treated as in orthodox Bayesianism. This model is interesting when it comes to representing ignorance or indifference: it can do so with a \emph{set of functions that disagree with each other.}

\subsection{Blurring Out Gott's Doomsday Argument: Apocalypse Not Now}
\label{DoomsdayIP}

\par Let us see how we can reframe Gott's Doomsday argument with an imprecise prior credence for the total number of humans $N$, or more generally for the length of any process of indefinite duration $X$. Let our prior credence in $X$, be represented by a family of credal functions $\{c_\gamma:c_\gamma\in C_X\}$, each normalizable and defined on $\mathbb{R}^{>0}$. Thus, we avoid improper prior distributions. All we assume is that $X$ is finite but can be indefinitely large. We have no reason to exclude from our prior credal set $C_X$ any distribution that is monotonically decreasing and such that $\forall c_\gamma\in C_X ,\lim_{X\rightarrow\infty}(c_\gamma(X))=0$.\footnote{In order to avoid too sharply peaked distributions (at $X\rightarrow0$), constraints can be placed on the variance of the distributions (namely, an lower bound on the variance), without it affecting my argument.} Let then our prior credence consist in the following set of functions, all of which decrease but not at the same rate (i.e., similar to a family of Pareto distributions), $\left\{c_\gamma(X)=\dfrac{k_\gamma}{X^\gamma}: c_\gamma\in C_X\right\},$ with $\gamma>1$ and $k_\gamma$ a normalizing constant: $k_\gamma=\dfrac{1}{\mathlarger{\int}_0^\infty\frac{\ud X}{X^\gamma}}$. The limiting case $\gamma\rightarrow1$ corresponds to $X\rightarrow\infty$, but $\gamma=1$ must be excluded to avoid a non-normalizable distribution.

If we don't want to assume anything about the distributions in $C_X$ (other than their being monotonically decreasing), this prior set must be such that it contains functions of decreasing rates that are arbitrarily small. That is, $\forall X\in\mathbb{R}^{>0}, \forall \epsilon\in\mathbb{R}^{<0}$, $\exists\,c_\gamma\in C_X$ such that $\dfrac{\ud c_\gamma(X)}{\ud X}>\epsilon$. This requirement applies not to any of the functions in $C_X$ but \emph{to the set as a whole}.

\par Following the steps of the argument given above in \S~\ref{Doomsday}, we obtain the following expression for the distributions in the credal set $\left\{c_\gamma(r):c_\gamma(r)\in C_r\right\}$ representing our prior credence in $r$:
\[
c_\gamma(r)=\int_{N=r}^{N=\infty}p(r|N)\cdot c_\gamma(N)\ud N
=\int_{N=r}^{N=\infty} \dfrac{k_\gamma}{N^{\gamma+1}} \ud N.
\]

\noindent Bayes' theorem then yields an expression for the posterior credal functions in $C_{N|r}$:
\[
c_\gamma(N|r)=\dfrac{p(r|N)\cdot c_\gamma(N)}{c_\gamma(r)}=\dfrac{k_\gamma}{N^{\gamma+1}\cdot \int_{N=r}^{N=\infty} \frac{k_\gamma}{N^{\gamma+1}}\ud N}.
\]

For each credal function in $C_{N|r}$, we can find a prediction for $N$ with a 95\%-level confidence, by solving $c_\gamma(N\leq x|r)=0.95$ for $x$, with
\[
c(N\leq x|r)=\mathlarger{\int}_r^x c(N|r)\ud N.
\]
We will find a prediction for $N$ given by our imprecise posterior credal set $C_{N|r}$ by determining its upper bound, i.e., a prediction all distributions in $C_{N|r}$ can agree on. Now, as $\gamma\rightarrow1$, the prediction for $x$ such that $c_\gamma(N\leq x|r)=95\%$ diverges. In other words, this imprecise representation of prior credence in $N$, reflecting our ignorance about $N$, does not yield any prediction about $N$.

Choosing any of the predictions given by the individual distributions in the credal set would be arbitrary. Without the possibility for my prior credence to be represented not by a single probability distribution but by an \emph{infinite set} of probability distributions, I cannot avoid obtaining an arbitrarily precise prediction. Other distributions (e.g., that decrease at different rates) could be added to the prior credal set, as long as they fulfill the criteria listed at the beginning of this section. However, no other distribution we could include would change this conclusion.

\subsection{Blurring Out Anthropic Predictions}

We are ignorant about what value of the vacuum energy density $\rho_V$ we should expect from our current theories. We can see that representing our prior ignorance or indifference about the value of the vacuum energy density $\rho_V$ by an imprecise credal set can limit, if not entirely dissolve, the appeal of anthropic considerations.

If we substitute imprecise prior and posterior credences in the formula from \citep[see infra \S~\ref{anthr}]{Weinberg2000}, we have $\ud C_{\rho_V}=\nu(\rho_V)\cdot C^\star_{\rho_V}\ud\rho_V,$ with $C^\star_{\rho_V}$ a prior credal set that will exclude all values of $\rho_V$ outside the anthropic range, and $\nu(\rho_V)$ the average number of galaxies which form for $\rho_V$, which as in \S~\ref{anthr} peaks around the mean value of the anthropic range. In order for the prior credence $C^\star_{\rho_V}$ to express our ignorance, it should be such that it doesn't favor any value of $\rho_V$.

\par With the imprecise model, such a state of ignorance can be expressed by a set of probability distributions $\{c^\star_i: c^\star_i\in C^\star_{\rho_V}\}$, all of which normalizable over the anthropic range and such that $\forall\rho_V, \exists\,c^\star_i, c^\star_j\in C^\star_{\rho_V}$ such that $\rho_V$ is favored by $c^\star_i$ and not by $c^\star_j$.\footnote{This can be obtained, for instance, by a family of Dirichlet distributions (preferable in order to have invariance under redescription \citep[see][]{DeCooman2009}), each of which giving an expected value at a different point in the anthropically allowed range. As in \S~\ref{DoomsdayIP}, a lower bound can be placed on the variance of all the functions in $C^\star_{\rho_V}$ in order to avoid dogmatic functions.} Such a prior credal set will not favor any value of $\rho_V$. In particular, it is possible to define this prior credal set so that for any value of $\rho_V$, the lowest expectation value among the the posteriors is lower than the highest expectation value among the priors. If then we adopt interval dominance as a criterion for decision-making (see infra \S~\ref{IP}), then no observation of $\rho_V$ will be able to lend support to our anthropic prediction.

One may object to the adoption of interval dominance in such a case. This criterion is arguably not fined-grained enough to help us for most of the inferences we are likely to encounter. However, this choice of demanding decision rule can be motivated by the fact that we have no plausible alternative theoretical framework to the anthropic argument. In this context, it can be reasonable to agree to increase one's credence about the anthropic explanation \emph{only if} it does better than any other yet unknown alternative might have done. Nonetheless, if we adopt other decision rules, it is possible with the imprecise model to construct prior credal sets that define a large interval over the anthropic range such that the confirmatory boost obtained after observing $\rho_V$ is not nearly as vindicative as it is with a single, uniform distribution.

This approach doesn't prevent Bayesian induction altogether. All the functions in $C^\star_{\rho_V}$ being probability distributions that can be treated as in orthodox Bayesianism; any of them can be updated and, in principle, converge toward a sharper credence, provided sufficient updating.

\section{Conclusion}

These cosmic puzzles show that, in the absence of an adequate representation of ignorance or indifference, a logic of induction will inevitably yield unwarranted results. Our usual methods of Bayesian induction are ill-equipped to allow us to address either puzzle. I have shown that the imprecise credence framework allows us to treat both arguments in a way that avoids their undesirable conclusions. The imprecise model rests on Bayesian methods, but it is expressively richer than the usual Bayesian approach that only deals with single probability distributions.

Philosophical discussions about the value of the imprecise model usually center around the difficulty to define updating rules that don't contradict general principles of conditionalization (especially the problem of dilation). But the ability to solve such paradoxes of confirmation and avoid unwarranted conclusions should be considered as a crucial feature of the imprecise model and play in its favor.



\begin{thebibliography}{}
\small{
\bibitem[\protect\citeauthoryear{Aguirre}{Aguirre}{2001}]{Aguirre2001}
Aguirre, Anthony. 2001.
\newblock {``Cold Big-Bang Cosmology as a Counterexample to Several Anthropic Arguments.''}
\newblock {\em Physical Review D\/}~64: 1--12.

\bibitem[\protect\citeauthoryear{Augustin, Coolen, de~Cooman, and Troffaes}{Augustin et~al.}{2014}]{Augustin2014}
Augustin, Thomas, Frank P.~A. Coolen, Gert de~Cooman, and Matthias~C.~M Troffaes (eds.). 2014.
\newblock {\em {Introduction to Imprecise Probabilities}}.
\newblock Wiley \& Sons.

\bibitem[\protect\citeauthoryear{Bostrom}{Bostrom}{2002}]{Bostrom2002}
Bostrom, Nick. 2002.
\newblock {\em {Anthropic Bias: Observation Selection Effects in Science and Philosophy}}.
\newblock New York: Routledge.

\bibitem[\protect\citeauthoryear{Carroll}{Carroll}{2000}]{Carroll2000}
Carroll, Sean.~M. 2000.
\newblock {``The Cosmological Constant.''}
\newblock {\em arXiv: astro-ph/0004075v2\/}, 1--50.

\bibitem[\protect\citeauthoryear{Dawid, Stone, and Zidek}{Dawid et~al.}{1973}]{Dawid1973}
Dawid, A.~Philip, M.~Stone, and James~V. Zidek. 1973.
\newblock {``Marginalization Paradoxes in Bayesian and Structural Inference.''}
\newblock {\em Journal of the Royal Statistical Society. Series B (Methodological)\/}~35: 189--233.

\bibitem[\protect\citeauthoryear{de~Cooman and Miranda}{de~Cooman and Miranda}{2007}]{DeCooman2007}
de~Cooman, Gert and Enrique Miranda. 2007.
\newblock {``Symmetry of Models Versus Models of Symmetry.''}
\newblock In William~L. Harper and Gregory Wheeler (eds.), {\em Probability and Inference. Essays in Honour of Henry E. Kyburg Jr}, 67--149. London: College Publications.

\bibitem[\protect\citeauthoryear{de~Cooman, Miranda, and Quaeghebeur}{de~Cooman et~al.}{2009}] {DeCooman2009}
de~Cooman, Gert, Enrique Miranda, and Erik Quaeghebeur. 2009.
\newblock {``Representation Insensitivity in Immediate Prediction under Exchangeability.''}
\newblock {\em International Journal of Approximate Reasoning\/}~50: 204--216.

\bibitem[\protect\citeauthoryear{Dieks}{Dieks}{1992}]{Dieks1992}
Dieks, Dennis. 1992.
\newblock {``Doomsday--Or: The Dangers of Statistics.''}
\newblock {\em The Philosophical Quarterly\/}~42: 78--84.

\bibitem[\protect\citeauthoryear{Goodman}{Goodman}{1994}]{Mackay1994}
Goodman, Steven~N. 1994.
\newblock {``Future Prospects Discussed.''}
\newblock {\em Nature\/}~368: 108--109.

\bibitem[\protect\citeauthoryear{Gott}{Gott}{1993}]{Gott1993}
Gott, J.~Richard. 1993.
\newblock {``Implications of the Copernican Principle for our Future Prospects.''}
\newblock {\em Nature\/}~363: 315--319.

\bibitem[\protect\citeauthoryear{Gott}{Gott}{1994}]{Gott1994}
--- --- --- 1994.
\newblock ``{Future Prospects Discussed.''}
\newblock {\em Nature\/}~368: 108.

\bibitem[\protect\citeauthoryear{Griffiths and Tenenbaum}{Griffiths and Tenenbaum}{2006}]{Griffiths2006}
Griffiths, Thomas~L. and Joshua~B. Tenenbaum. 2006.
\newblock {``Optimal Predictions in Everyday Cognition.''}
\newblock {\em Psychological Science\/}~17: 767--773.

\bibitem[\protect\citeauthoryear{Hartle and Srednicki}{Hartle and
  Srednicki}{2007}]{Hartle2007}
Hartle, James~B. and Mark Srednicki. 2007.
\newblock ``{Are We Typical?''}
\newblock {\em Physical Review D\/}~75: 123523--1--6.

\bibitem[\protect\citeauthoryear{Joyce}{Joyce}{2010}]{Joyce2010}
Joyce, James~M. 2010.
\newblock {``A Defense of Imprecise Credences in Inference and Decision Making.''}
\newblock {\em Philosophical Perspectives\/}~24: 281--323.

\bibitem[\protect\citeauthoryear{Leslie}{Leslie}{1990}]{Leslie1990}
Leslie, John~A. 1990.
\newblock {``Is the End of the World Nigh?''}
\newblock {\em The Philosophical Quarterly\/}~40: 65--72.

\bibitem[\protect\citeauthoryear{Levi}{Levi}{1974}]{Levi1974}
Levi, Isaac. 1974.
\newblock {On Indeterminate Probabilities.}
\newblock {\em The Journal of Philosophy\/}~71: 391--418.

\bibitem[\protect\citeauthoryear{Maor, Krauss, and Starkman}{Maor et~al.}{2008}]{Maor2008}
Maor, Irit, Lawrence Krauss, and Glenn Starkman. 2008.
\newblock {``Anthropic Arguments and the Cosmological Constant, with and without the Assumption of Typicality.''}
\newblock {\em Physical Review Letters\/}~100: 041301--1--4.

\bibitem[\protect\citeauthoryear{Martel, Shapiro, and Weinberg}{Martel et~al.}{1998}]{Martel1998a}
Martel, Hugo, Paul~R. Shapiro, and Steven Weinberg. 1998.
\newblock {``Likely Values of the Cosmological Constant}.''
\newblock {\em The Astrophysical Journal\/}~492: 29--40.

\bibitem[\protect\citeauthoryear{Monton and Roush}{Monton and Roush}{2001}]{Monton2001}
Monton, Bradley and Sherrilyn Roush. 2001.
\newblock ``{Gott's Doomsday Argument.''}
\newblock {\em http://philsci-archive.pitt.edu/id/eprint/1205\/}, 1--23.

\bibitem[\protect\citeauthoryear{Neal}{Neal}{2006}]{Neal2006}
Neal, Raford~M. 2006.
\newblock ``{Puzzles of Anthropic Reasoning Resolved Using Full Non-indexical Conditioning.''}
\newblock {\em Arxiv preprint math/0608592\/}~(0607), 1--56.

\bibitem[\protect\citeauthoryear{Norton}{Norton}{2010}]{Norton2010}
Norton, John~D. 2010.
\newblock {``Cosmic Confusions: Not Supporting versus Supporting Not.''}
\newblock {\em Philosophy of Science\/}~77: 501--523.

\bibitem[\protect\citeauthoryear{Pogosian, Vilenkin, and Tegmark}{Pogosian et~al.}{2004}]{Pogosian2004}
Pogosian, Levon, Alexander Vilenkin, and Max Tegmark. 2004.
\newblock {``Anthropic Predictions for Vacuum Energy and Neutrino Masses.''}
\newblock {\em Journal of Cosmology and Astroparticle Physics\/}~7: 1--17.

\bibitem[\protect\citeauthoryear{Richmond}{Richmond}{2006}]{Richmond2006}
Richmond, Alasdair. 2006.
\newblock {``The Doomsday Argument.''}
\newblock {\em Philosophical Books\/}~47: 129--142.

\bibitem[\protect\citeauthoryear{Sol\`{a}}{Sol\`{a}}{2013}]{Sola2013}
Sol\`{a}, Joan. 2013.
\newblock {``Cosmological Constant and Vacuum Energy: Old and New Ideas.''}
\newblock {\em Journal of Physics: Conference Series\/}~453: 012015--1--48.

\bibitem[\protect\citeauthoryear{Syversveen}{Syversveen}{1998}]{Syversveen1998}
Syversveen, Anne~Randi. 1998.
\newblock {Noninformative Bayesian Priors. Interpretation and Problems with Construction and Applications}. Unpublished manuscript.

\bibitem[\protect\citeauthoryear{Tegmark and Bostrom}{Tegmark and
  Bostrom}{2005}]{Tegmark2005}
Tegmark, Max and Nick Bostrom. 2005.
\newblock {``Is a Doomsday Catastrophe Likely?''}
\newblock {\em Nature\/}~438: 754.

\bibitem[\protect\citeauthoryear{Troffaes}{Troffaes}{2007}]{Troffaes2007}
Troffaes, Matthias~C. 2007.
\newblock {``Decision Making under Uncertainty Using Imprecise Probabilities.''}
\newblock {\em International Journal of Approximate Reasoning\/}~45: 17--29.

\bibitem[\protect\citeauthoryear{Vilenkin}{Vilenkin}{1995}]{Vilenkin1995}
Vilenkin, Alexander. 1995.
\newblock {``Predictions from Quantum Cosmology.''}
\newblock {\em Physical Review Letters\/}~74: 4--7.

\bibitem[\protect\citeauthoryear{Walley}{Walley}{1991}]{Walley1991}
Walley, Peter. 1991.
\newblock {\em {Statistical Reasoning with Imprecise Probabilities}}.
\newblock London: Chapman and Hall.

\bibitem[\protect\citeauthoryear{Weinberg}{Weinberg}{1987}]{Weinberg1987}
Weinberg, Steven. 1987.
\newblock {``Anthropic Bound on the Cosmological Constant.''}
\newblock {\em Physical Review Letters\/}~59: 2607--2610.

\bibitem[\protect\citeauthoryear{Weinberg}{Weinberg}{2000}]{Weinberg2000}
--- --- --- 2000.
\newblock {``A Priori Probability Distribution of the Cosmological Constant.''}
\newblock {\em arXiv preprint \href{http://arxiv.org/abs/astro-ph/0002387}{astro-ph/0002387}\/}, 0--15.

\bibitem[\protect\citeauthoryear{Weinberg}{Weinberg}{2007}]{Weinberg2007}
--- --- --- 2007.
\newblock {``Living in the Multiverse.''}
\newblock In Bernard Carr (ed.). {\em Universe or Multiverse?}, Chapter~2, 29--42. Cambridge: Cambridge University Press.

\bibitem[\protect\citeauthoryear{Wells}{Wells}{2009}]{Wells2009}
Wells, Willard. 2009.
\newblock {\em {Apocalypse When? Calculating How Long the Human Race Will Survive}}.
\newblock Chichester, UK: Springer Praxis Books. Praxis.
}
\end{thebibliography}
\end{document}